\documentclass[prl,twocolumn,showpacs,amssymb,superscriptaddress]{revtex4-1}
\usepackage{amsmath}
\usepackage{graphicx}% Include figure files
\usepackage{epsfig}
\usepackage{dcolumn}% Align table columns on decimal point
\usepackage{bm}% bold math

\DeclareMathOperator{\re}{Re} % NB: `\Re' is already taken! 
\DeclareMathOperator{\im}{Im} % NB: `\Im' is already taken!

\usepackage{epsfig}
\def\gsim{\lower0.5ex\hbox{$\:\buildrel >\over\sim\:$}}
\def\lsim{\lower0.5ex\hbox{$\:\buildrel <\over\sim\:$}}
\newcommand{\bea}{\begin{eqnarray}}
\newcommand{\eea}{\end{eqnarray}}

\begin{document}
%\thispagestyle{empty}
%\vspace*{0.01cm}
%\vspace{.01cm}
\vspace{-0.2cm}
\begin{flushright}
\hspace*{5.5in}
\mbox{HIP-2010-07/TH}
\end{flushright}
\title{Ultrarelativistic sneutrinos at the LHC and sneutrino-antisneutrino oscillation}
\author{Dilip Kumar Ghosh}%
\email{tpdkg@iacs.res.in}
\affiliation{Department of Theoretical Physics and Centre for Theoretical
Sciences, Indian Association for the Cultivation of Science, 2A $\&$ 2B Raja
S.C. Mullick Road, Kolkata 700 032, India}
\author{Tuomas Honkavaara}%
\email{Tuomas.Honkavaara@helsinki.fi}
\affiliation{Department of Physics, and Helsinki Institute of Physics,
P.O. Box 64, FIN-00014 University of Helsinki, Finland}
\author{Katri Huitu}%
\email{Katri.Huitu@helsinki.fi}
\affiliation{Department of Physics, and Helsinki Institute of Physics,
P.O. Box 64, FIN-00014 University of Helsinki, Finland}
\author{Sourov Roy}%
\email{tpsr@iacs.res.in}
\affiliation{Department of Theoretical Physics and Centre for Theoretical
Sciences, Indian Association for the Cultivation of Science, 2A $\&$ 2B Raja
S.C. Mullick Road, Kolkata 700 032, India}
\date{\today}
\begin{abstract}
Sneutrino-antisneutrino oscillation can be a very useful probe to look 
for signatures of lepton number violation ($\Delta L$ = 2) at the LHC. 
Here, we discuss the effect of the Lorentz factor $\gamma$ and the travelling 
distance $L$ on the probability of the oscillation. We demonstrate that these 
two parameters can significantly alter the probability of the oscillation when 
the sneutrinos are ultrarelativistic and have a very small total decay width. 
We propose a scenario where these requirements are fulfilled and which produces 
interesting signals at the LHC even for a mass splitting $\Delta m$ as small as 
$10^{-14}$ GeV between the sneutrino mass eigenstates.
\end{abstract}


\maketitle

%------------------------------------------------------------------------------
\newpage

Oscillation in neutral systems, like $K^0$, $B^0_d$, $B^0_s$, $D^0$, and
$\nu$, has been measured and has provided important understanding
of weak interactions and, in the case of neutrinos, revealed their
nonzero masses. If the neutrino mass is of the Majorana type, it is expected 
that the supersymmetric partners of neutrinos, i.e., sneutrinos, oscillate 
analogously to the neutral meson system. A major difference between neutral 
meson and neutrino oscillations is that, in the meson oscillations, similarly to 
the sneutrino oscillations, the decay of the oscillating particle has to be taken
into account. In most applications so far, the neutral meson can be considered a 
nonrelativistic particle, e.g., this is true in $B$-factories. The oscillation of 
neutrinos is obviously between relativistic particles, but the system is qualitatively 
different otherwise, since the neutrinos do not decay.

Sneutrino-antisneutrino oscillation probes the lepton number violation
($\Delta L =2$) and can be present when the neutrinos 
have nonzero Majorana masses 
\cite{hirschetal, grossman-haber, choietal, chun, tuomas-egamma, dedes-haber-rosiek,tuomas-lhc}.
It can also provide information on the neutrino sector parameters at the
collider environment \cite{choietal, tuomas-egamma}.
However, in the derivation of the sneutrino-antisneutrino oscillation 
probability, one usually assumes that the sneutrinos are produced at rest, 
as in the case of $K^0$--${\bar{K}^0}$ oscillation or 
$B^0$--${\bar{B}^0}$ oscillation. % \cite{branco-etalii}. 
The situation is different
when we produce sneutrinos at the LHC energy, and it is not correct to assume that they are produced at rest. 

In this paper, we outline the calculation of a formula for 
the sneutrino-antisneutrino oscillation probability that is 
applicable when the sneutrino is produced with a very high energy and momentum, as is the case, e.g., at the 
LHC. In the studies of sneutrino-antisneutrino oscillation so far, only the nonrelativistic case has been 
considered. We also stress the importance of using the correct formula in the context of a very interesting 
supersymmetric scenario which can produce spectacular signals at the LHC.

Let us first write down the sneutrino (${\tilde \nu}$) and antisneutrino
(${\tilde \nu}^*$) states in terms of the mass eigenstates,
\bea
|{\tilde \nu}\rangle = \frac{1}{\sqrt{2}}(|{\tilde \nu}_1 \rangle + 
i |{\tilde \nu}_2 \rangle), ~ |{\tilde \nu}^*\rangle = 
\frac{1}{\sqrt{2}}(|{\tilde \nu}_1 \rangle - i |{\tilde \nu}_2 \rangle). \nonumber \\
\label{sneu_antisneu_state}
\eea
The state $|\tilde\nu\rangle $ at $(x,t)$ becomes 
\bea
|\psi(x,t)\rangle = \frac{1}{\sqrt{2}} \left[e^{-i(Et -{p_1}{x})} 
|{\tilde \nu}_1 \rangle +  ie^{-i(Et -{p_2}{x})} |{\tilde \nu}_2 \rangle \right]. 
\hspace*{-0.6cm}\nonumber \\
\label{sneu_gen_state}
\eea
The mass eigenstates are $|{\tilde \nu}_1 \rangle$ and 
$|{\tilde \nu}_2 \rangle$ with three-momenta $p_1$ and $p_2$, respectively.
Here, we assume that the mass eigenstates move with the 
same energy $E$ but different three-momenta $p_1$ and $p_2$. 

The probability of a $|{\tilde \nu}\rangle$ oscillating into 
an $|{\tilde \nu}^* \rangle$ is then given by 
\bea
P_{{\tilde \nu}\to{\tilde \nu}^*} = |\langle{\tilde \nu}^*|\psi(x,t)\rangle|^2.
\eea
Using Eqs. (\ref{sneu_antisneu_state}) and (\ref{sneu_gen_state}), 
we can expand the probability as
\bea
P_{{\tilde \nu}\to{\tilde \nu}^*} &=& \frac{1}{4} \Big[e^{-2\im(p_1)x} + e^{-2\im(p_2) x}
\nonumber \\
& & -e^{i(p_2 - p_1^*)x} - e^{i(p_1 - p_2^*)x} \Big].
\label{eq_osc_prob}
\eea
Including the effect of the total decay widths of the sneutrino mass 
eigenstates, one can write down the three-momenta $p_i$, with $i=1,2$, as $p_i 
= \sqrt{E^2 - m^2_i + i \Gamma m_i}$. Here, we assume that the total decay widths are the same for 
the sneutrino and the antisneutrino, and the width is denoted by 
$\Gamma$. In addition, $m_1$ and $m_2$ are the mass eigenvalues 
of the sneutrino mass eigenstates 
$|{\tilde \nu}_1 \rangle$ and $|{\tilde \nu}_2 \rangle$, respectively. 
Assuming a very small $\Gamma $ and 
%that the sneutrinos are 
%highly relativistic, i.e. $E \gg \Gamma, m_1, m_2$, we can 
$E \gg \Gamma, m_1, m_2$, we can 
approximate $p_i$, with $i=1,2$, as 
\bea
p_i &\simeq& \sqrt{E^2 -m^2_i} \left[1 + \frac{i \Gamma m_i}{2(E^2 -m^2_i)} 
\right].
\eea
The last two terms of Eq.~(\ref{eq_osc_prob}) can be 
expanded 
using standard trigonometric and hyperbolic formulae with 
$ \im(p_i) \simeq \frac{\Gamma m_i}{2E} $ and 
$ \re(p_i) \simeq E - \frac{m^2_i}{2E} $.
Hence, we can calculate the probability of a $|{\tilde \nu}\rangle$ oscillating into an 
$|{\tilde \nu}^*\rangle$ as
\bea
P_{{\tilde \nu}\to{\tilde \nu}^*} &=& 
|\langle {\tilde \nu}^*|\psi(x,t)\rangle|^2 = \frac{1}{4}
\bigg[e^{-{\frac{\Gamma m_1}{E}}x} + e^{-{\frac{\Gamma m_2}{E}}x} 
\nonumber \\
& & -2\cos\left(\frac{\Delta m^2}{2E} x\right) e^{-{\frac{\Gamma}{2E}}(m_1 + m_2)x} 
\bigg]. 
\label{sneu_osc_exponential_suppression}
\eea
Here, $\Delta m^2 \equiv m^2_1 -m^2_2$.
In the appropriate limit, this formula agrees with the formula for
neutral meson mixing with very large momenta \cite{Burgess:2007zi}.

Since the sneutrinos (antisneutrinos) decay, we need to look at the
integrated probability. 
Assuming $m_1 \approx m_2 = m$, we get 
$ \langle \psi(x,t)|\psi(x,t)\rangle \simeq e^{-{\frac{\Gamma m}{E}}x} $
and 
$ \int_0^\infty dx \langle \psi(x,t)|\psi(x,t)\rangle = \frac{E}{\Gamma m} $.
The integrated probability, at a distance $L$, of a $|\tilde \nu\rangle$ oscillating 
into an $|{\tilde \nu}^*\rangle$ is given by
\bea
P(L) &=& \frac{\int_0^L dx |\langle{\tilde \nu}^*|\psi(x,t)\rangle|^2}
{\int_0^\infty dx \langle \psi(x,t)|\psi(x,t)\rangle} \nonumber \\
&=& \frac{e^{-L\alpha}}{2 (\alpha^2 + \beta^2)}\Big[-\alpha^2 + (-1 + e^{L\alpha})\beta^2 
\nonumber \\
& & +\alpha^2 \cos(L\beta) - \alpha \beta \sin(L\beta)\Big], 
\label{length_dependent_int_osc_formula}
\eea
where $\alpha \equiv \frac{\Gamma m}{E}$ and $\beta \equiv \frac{\Delta m^2}{2E}$. 
For a very large $L$, i.e., when $L \alpha \gg$ 1, from 
Eq. (\ref{length_dependent_int_osc_formula}), we get
\bea
P(L) = \frac{\beta^2}{2 (\alpha^2 + \beta^2)} = 
\frac{x^2_{\tilde \nu}}{2(1 + x^2_{\tilde \nu})},
 \label{old_osc_prob}
\eea
which is independent of $L$ and
where we use the relation $\Delta m^2 = 2m \Delta m$ and 
$x_{\tilde \nu}$ is defined as $x_{\tilde \nu} \equiv \frac{\Delta m}{\Gamma}$ \cite{grossman-haber}.
Equation (\ref{old_osc_prob}) is the same result as in the case when the sneutrinos 
are produced at rest. 
Note from Eq. (\ref{old_osc_prob}) that, with $L\alpha \gg$ 1, when 
$x_{\tilde \nu}$ = 1, the oscillation 
probability $P({\tilde \nu} \rightarrow {\tilde \nu}^*)$ is 0.25. On the other 
hand, when 
$x_{\tilde \nu} \gg$ 1, $P({\tilde \nu} \rightarrow {\tilde \nu}^*)$ is 0.5. 
We can see from Eq. (\ref{sneu_osc_exponential_suppression}) that the 
oscillation probability has an exponential suppression factor.  

\begin{figure}
\includegraphics[scale=0.6]{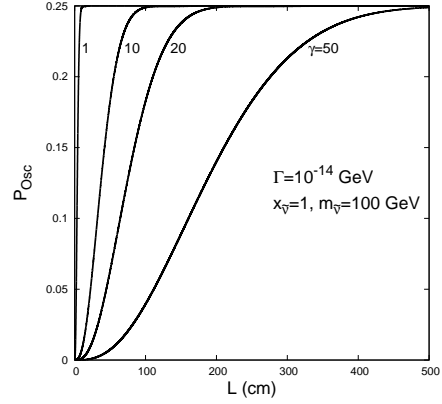}
\caption{Dependence of the integrated sneutrino oscillation probability on the Lorentz factor $\gamma$ when $\frac{\Delta m}{\Gamma} = 1$ and $\Gamma = 10^{-14}$ GeV. Note that the case of $\gamma=1$ is actually calculated with $\gamma=1.0001$.}
\label{osc_prob_old_boost_effect}
\end{figure}

Next, let us investigate what the effect of the Lorentz factor 
$\gamma = \frac{E}{m}$ on the sneutrino 
oscillation probability is. In order to do this, we must keep the length 
dependence of the oscillation
probability formula (see Eq. (\ref{length_dependent_int_osc_formula})). 
Hence, we consider 
$L\alpha \lsim \mathcal{O}(1)$. Note that the quantity 
$\alpha \equiv \frac{\Gamma m}{E} 
= \frac{\Gamma}{\gamma}$ is the sneutrino (antisneutrino) decay width modified 
by the Lorentz factor.   
In Fig. \ref{osc_prob_old_boost_effect}, we plot the integrated sneutrino oscillation probability 
$P({\tilde \nu} \rightarrow {\tilde \nu}^*)$ as a function of the travelling 
distance $L$.
The four different lines on this plot correspond to four different values of 
the Lorentz factor $\gamma$ of the produced sneutrino. We assume that the total 
decay width ($\Gamma$) of a 100 GeV mass sneutrino is $10^{-14}$ GeV and $x_{\tilde \nu} = 1$.  
It is seen from this plot that the oscillation probability has a strong dependence on $\gamma$ up to 
a certain value of $L$, and, after that, it saturates and reaches the value 0.25, independent of $\gamma$ 
and $L$. As long as the $L$-dependence is there, for a particular value of $L$, 
$P({\tilde \nu} \rightarrow {\tilde \nu}^*)$ is smaller for a larger value of $\gamma$. 
This can be understood as follows. Looking at Eq. (\ref{sneu_osc_exponential_suppression}), we see that 
there is a length-dependent exponential suppression factor that also depends on $\gamma =E/m$. For smaller 
values of $\gamma$, this produces a sharper variation of the oscillation probability; whereas, for larger values 
of $\gamma$, the variation is relatively slow. This is also reflected in the variation of the integrated 
oscillation probability with distance, see Fig. \ref{osc_prob_old_boost_effect}.

It is, however, interesting to note that, for a much higher value of $\Gamma (\sim 10^{-7} {\rm GeV })$ 
(with $x_{\tilde{\nu}} \gtrsim 1$), 
the value of $L$ is very small ($\approx 5 \times 10^{-5}$ cm) for which the oscillation 
probability saturates 
(even for $\gamma$ = 50). Hence, for such a large value of $\Gamma $,
we can ignore the effect of
$\gamma$ or $L$ in the sneutrino oscillation probability. 

On the other hand, if the sneutrino (antisneutrino) decay width $\Gamma$ is 
much smaller 
(i.e., $\sim 10^{-14}$ GeV or so), 
the $L$- and $\gamma$-dependences are 
much more pronounced.
In such a situation, one should use 
the probability formula given in Eq. (\ref{length_dependent_int_osc_formula}). 
Such small values of the sneutrino
decay width are possible, 
for example, in a scenario where the
left-handed sneutrino NLSP is 
nearly degenerate to the lighter stau LSP and the dominant decay channel for ${\tilde \nu}_\tau$ is 
\bea
{\tilde \nu}_\tau \rightarrow {\tilde \tau}_1^- + \pi^+,
\eea
with a total decay width $\Gamma \sim 10^{-14}$ GeV. In some models with an
extra $U(1)_{B-L}$, the oscillation 
of a right-chiral sneutrino (${\tilde \nu}_R$) can be important \cite{khalil_B-L_RHSNU_osc}. In such cases, the 
total decay width of ${\tilde \nu}_R$ can be as small as $\sim 10^{-14}$ GeV. 
The left-chiral sneutrino decay 
width can also be reduced if it has a significant mixing with the right-chiral
counterpart.

When the dominant sneutrino decay is ${\tilde \nu}_\tau \rightarrow {\tilde \tau}_1^- + \pi^+$,
one can see a signal 
$pp \rightarrow {\tilde \nu}_\tau {\tilde \tau}_1^+ \rightarrow {\tilde \tau}_1^- {\tilde \tau}_1^+ + \pi^+$. 
This produces two heavily ionized charged tracks with opposite curvatures when there is no oscillation and with same 
curvatures when there is sneutrino oscillation. We assume that these 
stau tracks can be distinguished from the muon 
tracks, due to the slower velocity of staus. 
Similarly, one should also look at the signal 
$pp \rightarrow {\tilde \nu}^*_\tau {\tilde \tau}_1^- \rightarrow {\tilde \tau}_1^+ {\tilde \tau}_1^- + \pi^-$. 
In this case, when the sneutrino oscillates, one can see two same-sign heavily ionized charged tracks due to a 
pair of ${\tilde \tau}_1^-$s. Note that the sneutrino is long-lived (decay length approximately a few centimeters), and, hence, 
one of the staus produced from the decay of the sneutrino shows a secondary vertex which is well separated from 
the primary vertex. This is a very spectacular signal and free from any standard model (SM) or supersymmetric (SUSY) backgrounds. This parameter region provides 
distinct phenomenology, and one might consider taking a point in the region as a benchmark point for a general minimal supersymmetric standard model (MSSM).

In order to get an idea about the cross section and the branching ratio of the processes
discussed, we consider a mass spectrum with a ${\tilde \nu}_\tau$ as the next-to-lightest supersymmetric particle (NLSP) and a ${\tilde \tau}_1$ as the lightest supersymmetric particle (LSP).
We include a tiny $R$-parity violating ($RPV$) coupling such that the ${\tilde \tau}_1$ decays outside the detector, leaving a 
heavily ionized charged track. We assume that this small $RPV$ coupling does not change the total decay
width of the sneutrino. 

Nevertheless, below, we list different regions of interest for the strength of the $RPV$ coupling 
and the width $\Gamma$. In all cases, we assume 
$x_{\tilde\nu}=\Delta m/\Gamma \gsim 1$  (for $x_{\tilde\nu}=0.2$, the maximum oscillation probability drops to 2\%).
\begin{enumerate}

\item The $RPV$ coupling is very small ($\lsim 10^{-8}$) and 
$\Gamma \lsim 10^{-13}$ GeV, as considered in this analysis.  In this case, the effect of our Eq. (7) is prominent.

\item The $RPV$ coupling is larger than what we consider but does not increase the sneutrino total decay width 
significantly ($10^{-14}$ GeV $< \Gamma <$ $10^{-13}$ GeV), and the $RPV$ coupling is $\sim 10^{-7}$. For such a value of 
the $RPV$ coupling, the stau may decay inside the detector, leaving a heavily ionizing charged track with a kink. In this case,
the displaced vertex ($\gsim$ a few mm) from the sneutrino will be present. 

\item The sneutrino total decay width is larger (but $\Gamma \lsim 10^{-7}$ GeV) and the $RPV$ coupling  is small, 
$\lsim 10^{-8}$. In this case, the sneutrino oscillation signals remain with charged tracks from the long-lived stau. However, 
in this case, the effect of the boost and the displaced vertex from the sneutrino will be absent \cite{tuomas-lhc}.

\item $\Gamma \lsim 10^{-7}$ GeV but the $RPV$ coupling is larger, $\gsim 10^{-7}$.  Both the displaced vertex and the stau 
track will be absent, and one has to worry about the SM/SUSY backgrounds. 
\end{enumerate}

The mass of the sneutrino is considered 
to be $m_{{\tilde \nu}_\tau} = 100$ GeV and the mass of
${\tilde \tau}_1$ is $m_{{\tilde \tau}_1} = 99.7$ GeV. The stau mixing angle is taken to be $\pi/4$. The other relevant 
parameter choices are $M_1 = 120$ GeV, $M_2 = 240$ GeV, $\mu = -250$ GeV, $\tan\beta=6$, $m_{A^0}=600$ GeV and 
$A_\tau = 250$ GeV. Here, $M_1$ and $M_2$ are the $U(1)$ and $SU(2)$ gaugino mass parameters, respectively, $\mu$ is the 
superpotential $\mu$-parameter, $m_{A^0}$ is the pseudoscalar Higgs boson mass and $A_\tau$ is the trilinear scalar coupling 
of the staus. 
With these values of parameters,
%For such a choice of the mass spectrum, 
the total decay width of the sneutrino is $\Gamma \approx 1 \times 
10^{-14}$ GeV,
while
the branching ratio of the decay ${\tilde \nu}_\tau \rightarrow {\tilde \tau}_1^- + \pi^+$ 
is 93$\%$. In fact, the branching ratio is greater than 90$\%$ when the mass splitting between the ${\tilde \nu}_\tau$ and 
the ${\tilde \tau}_1$ is in the range 200--350 MeV. Let us then consider the production
cross sections at the LHC. We get 
the opposite-sign (OS) stau signal  
$pp \rightarrow {\tilde \tau}_1^+ {\tilde \tau}_1^-$ 
from both ${\tilde \nu}_\tau 
{\tilde \tau}_1^+$ and ${\tilde \nu}^*_\tau {\tilde \tau}_1^-$ 
productions with an {\it effective} survival probability $(1-P_\mathrm{eff})$.
%On the other hand, 
The same-sign (SS) stau signal  
$pp \rightarrow {\tilde \tau}_1^+ {\tilde \tau}_1^+$ or ${\tilde \tau}_1^- 
{\tilde \tau}_1^-$ 
we get 
from either ${\tilde \nu}_\tau {\tilde \tau}_1^+$ 
or ${\tilde \nu}^*_\tau {\tilde \tau}_1^-$ productions with the {\it effective} oscillation
probability $(P_\mathrm{eff})$. 

We select the signal events with the following criteria: 
1) the pseudorapidities of the staus 
must be $|\eta^{{\tilde \tau}_1}|< 2.5$, 
2) the isolation variable 
$\Delta R \equiv \sqrt{(\Delta \eta)^2 + (\Delta \phi)^2}$ 
should satisfy $\Delta R> 0.7$ 
for the two staus, 
3) the transverse momentum of both staus must satisfy 
$p^{{\tilde \tau}_1}_T> 20$ GeV and 
4) the $\beta\gamma$ 
should be $0.3 < \beta\gamma < 2.0$. 
The upper limit of $\beta\gamma$ reduces the 
muon background considerably. Applying these cuts, the cross sections
with different center of mass energies and different 
$\Delta m$ are presented in Table \ref{tab_cross_sec} for $L=0.10$ m. From Table \ref{tab_cross_sec}, it is clear
that, for $\Delta m \gtrsim 10^{-13}$ GeV, the cross sections almost saturate.
Even putting $\Delta m$ to its maximum value, 
$10^{-7}$ GeV (see Eq. (8) of Ref. \cite{grossman-haber}),
does not change the
cross sections from $\Delta m = 10^{-10}$ GeV values. 
On the other hand, we can probe 
down to $\Delta m = 10^{-14}$ GeV 
and 
measure several SS events even with 10 fb$^{-1}$ luminosity.
%with 100 fb$^{-1}$ luminosity (or even with 10 fb$^{-1}$).
\begin{table}
\begin{center}
\footnotesize
\begin{tabular}{|c|c|c|c|c|c|c|}
\hline
$\Delta m$ [GeV] & \multicolumn{2}{c|}{$10^{-14}$} & \multicolumn{2}{c|}{$10^{-13}$} & \multicolumn{2}{c|}{$10^{-10}$} \\
\hline
 & \multicolumn{6}{c|}{Cross section in fb} \\
\hline
Signal & OS & SS & OS & SS & OS & SS \\
\hline
$\sqrt{s}=7$ TeV & 31.0 & 8.1 & 20.6 & 18.6 & 20.3 & 18.8 \\
\hline
$\sqrt{s}=12$ TeV & 52.0 & 13.6 & 34.4 & 31.2 & 34.1 & 31.6 \\
\hline
$\sqrt{s}=14$ TeV & 60.2 & 15.8 & 39.9 & 36.1 & 39.4 & 36.5 \\
\hline
\end{tabular}
\end{center}
\caption{Cross sections for the OS and SS stau signals with several center of mass energies and $\Delta m$. Here, $L=0.10$ m. The cuts used are mentioned in the text.}
\label{tab_cross_sec}
\end{table}
Using the SS and OS cross sections, we define the asymmetry
$ A = \frac{\sigma({\rm SS}) - \sigma({\rm OS})}
{\sigma({\rm SS}) + \sigma({\rm OS})}$.
For the purpose of illustration, we show one value of this asymmetry,
$A = -0.038\pm 0.011$, obtained by using 
$\sigma_{\rm SS}$ and $\sigma_{\rm OS}$
for $\sqrt{s} = 14 $ TeV with $\Delta m = 10^{-10}$ GeV from Table \ref{tab_cross_sec} and
assuming an integrated luminosity of $100~{\rm fb}^{-1}$. 
This asymmetry $A$ gives direct information about the oscillation 
probability and is independent of initial state parton densities and other 
uncertainities arising from higher order corrections. 
It is easy to check that $P_{\rm eff} = (1+A)/2$. 
By measuring the value of $A$, one can 
%easily 
calculate the 
{\it effective} oscillation probability. 
For our example, we get $P_{\rm eff} = 0.48$.

In Table \ref{tab_cross_sec_2}, there are the cross sections
with different center of mass energies and different 
$\Delta m$ for $L=0.30$ m. These $L=0.30$ m values already correspond to the nonrelativistic oscillation probability (i.e., we have Eq. (\ref{old_osc_prob}) at hand). This means that, for example, the SS values for $\sqrt{s}=14$ TeV in Table \ref{tab_cross_sec} become $\sim 4-14$\% higher if the nonrelativistic formula is used.
\begin{table}
\begin{center}
\footnotesize
\begin{tabular}{|c|c|c|c|c|c|c|}
\hline
$\Delta m$ [GeV] & \multicolumn{2}{c|}{$10^{-14}$} & \multicolumn{2}{c|}{$10^{-13}$} & \multicolumn{2}{c|}{$10^{-10}$} \\
\hline
 & \multicolumn{6}{c|}{Cross section in fb} \\
\hline
Signal & OS & SS & OS & SS & OS & SS \\
\hline
$\sqrt{s}=7$ TeV & 29.9 & 9.3 & 19.8 & 19.3 & 19.6 & 19.6 \\
\hline
$\sqrt{s}=12$ TeV & 50.1 & 15.6 & 33.2 & 32.5 & 32.8 & 32.8 \\
\hline
$\sqrt{s}=14$ TeV & 57.9 & 18.0 & 38.4 & 37.6 & 38.0 & 38.0 \\
\hline
\end{tabular}
\end{center}
\caption{Cross sections for the OS and SS stau signals with several center of mass energies and $\Delta m$. Here, $L=0.30$ m. The cuts used are mentioned in the text.}
\label{tab_cross_sec_2}
\end{table}

\begin{figure}
\begin{center}
\includegraphics[scale=0.35]{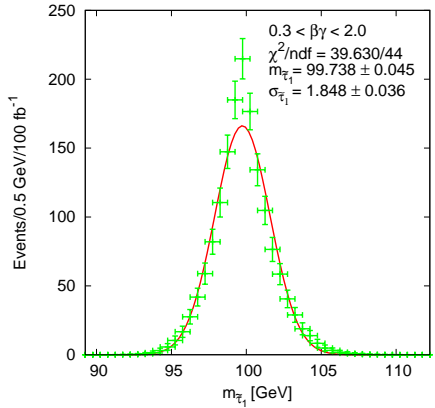}
\caption{The measured stau mass from the SS with $\sqrt{s}=14$ TeV, $\Delta m=10^{-14}$ GeV, and $L=0.10$ m. The cuts used are mentioned in the text.} 
\label{stau_mass_fit}
\end{center}
\end{figure}

If one can measure 
%the magnitude of 
the three-momentum ($|\vec p|$) of the stau
track and the corresponding $\beta \gamma$ at the LHC, then one can get 
an estimate of the stau mass $ m_{{\tilde \tau}_1} = \frac{|\vec p|}
{\beta \gamma} $ \cite{ellis-raklev-oye,ibe-kitano}. 
The plot of the measured stau mass coming from the SS with $\sqrt{s}=14$ TeV, $\Delta m=10^{-14}$ GeV, and $L=0.10$ m is shown in Fig. \ref{stau_mass_fit}.  All the cuts mentioned in an earlier paragraph 
are used here. The stau momentum and the velocities are smeared according to the 
formulae given in Ref. \cite{ibe-kitano}. 
The mass of the decaying sneutrino can be measured from the transverse mass 
distribution of the sneutrino.

In conclusion, sneutrino oscillation is a very important tool to look for lepton number violation at the LHC. 
However, at the LHC, the sneutrino can be ultrarelativistic, and one should appropriately take into account 
the Lorentz factor $\gamma \equiv \frac{E}{m}$ and the $L$-dependence while calculating the probability of 
oscillation. We have seen that the effect is more pronounced when the total decay width of the sneutrino is 
very small ($\sim 10^{-14}$ GeV), and this can be realized in many different SUSY scenarios. A very interesting 
signal at the LHC could be two same-sign heavily ionized charged tracks and a soft pion, which can probe a 
mass splitting all the way down to $\sim 10^{-14}$ GeV with an integrated luminosity as low as 10 ${\rm fb}^{-1}$ %\\[3pt] 
for $\sqrt{s}$ = 14 TeV. In fact, for the same mass splitting, it is very evident from Tables I and II that, even 
for $\sqrt{s}$ = 7 TeV with an integrated luminosity as low as 0.5--1 ${\rm fb}^{-1}$, one would expect to see 4--8 sneutrino 
oscillation events.

%\begin{flushleft}
%{\large \bf Acknowledgements}
%\end{flushleft}

We thank P. Eerola, P. Ghosh, M. Maity, P. Majumdar and B. Mukhopadhyaya for discussions.
This work is supported in part by the Academy of Finland (Project No. 115032).
D.K.G. acknowledges partial support from the Department of Science and Technology, India, under the grant 
SR/S2/HEP-12/2006. T.H. thanks the V\"ais\"al\"a Foundation for support.

\end{document}